# Predictive Crypto-Asset Automated Market Making Architecture for Decentralized Finance using Deep Reinforcement Learning

Tristan LIM*

*School of Business Management, Nanyang Polytechnic, Singapore*

*School of Computing and Information Systems, Singapore Management University, Singapore*

*tristan_lim@nyp.edu.sg; tristanl.2021@smu.edu.sg; tris02@gmail.com

**Abstract**

The study proposes a quote-driven predictive automated market maker (AMM) platform with on-chain custody and settlement functions, alongside off-chain predictive reinforcement learning capabilities to improve liquidity provision of real-world AMMs. The proposed AMM architecture is an augmentation to the Uniswap V3, a cryptocurrency AMM protocol, by utilizing a novel market equilibrium pricing for reduced divergence and slippage loss. Further, the proposed architecture involves a predictive AMM capability, utilizing a deep hybrid Long Short-Term Memory (LSTM) and Q-learning reinforcement learning framework that looks to improve market efficiency through better forecasts of liquidity concentration ranges, so liquidity starts moving to expected concentration ranges, prior to asset price movement, so that liquidity utilization is improved. The augmented protocol framework is expected have practical real-world implications, by (i) reducing divergence loss for liquidity providers, (ii) reducing slippage for crypto-asset traders, while (iii) improving capital efficiency for liquidity provision for the AMM protocol. To our best knowledge, there are no known protocol or literature that are proposing similar deep learning-augmented AMM that achieves similar capital efficiency and loss minimization objectives for practical real-world applications.

## Introduction

The introduction of smart contracts, backed by public blockchains such as Ethereum, allowed the creation of an entire financial system where different parties can operate under shared data and assumptions without trust issues arising from institutional intervention. This is also known as decentralised finance (DeFi).

DEX represents an important element of the DeFi market structure. Prior to the advent of decentralized exchanges (DEX) in recent years, trading of blockchain-derivative assets are generally conducted on off-chain, centralized settlement infrastructure. These off-chain order-driven exchanges, also known as centralized exchanges (CEX), act as trusted third parties. Examples of CEX are Binance and Bitfinex. While CEX offers easy-to-understand order book format execution similar to conventional financial market exchanges, it can experience server downtime, uncertain fair execution, slow withdrawals, and traders are wholly dependent on trust with the exchange on their custody of assets. Over time, there exist semi-custodial exchanges that seeks to move partial functionality on-chain. Examples of such exchanges are EtherDelta and IDEX, which deploy an on-chain custody and settlement solution, with an off-chain order book and trading engine. While the original intent is to create improved performance, downsides of CEX persist.

A new class of quote-driven crypto-asset trade execution system was developed, that requires only data structures and traversals, with low gas complexity (Moosavi and Clark, 2021). Known as automated market makers (AMM), these market making systems allow multiple parties to interact directly in a non-rivalrous and programmatic manner with smart contracts of the DEX protocol, so that trade executes automatically using a hard-coded pricing function (or a bonding curve), and matching of individual buy and sell orders are not required. Lehar and Parlour (2021) provided evidence of uptake of liquidity sharing AMM protocols and demonstrated empirically that AMM can provide liquidity more efficiently than CEX.

Market participants of an AMM are as follows:

- *Liquidity taker*:

  A liquidity taker is any party that exchanges assets by taking liquidity from the market, supplied by liquidity makers. They expect the market to reflect true price of assets, low price change during trade execution (or slippage), and the capacity to exchange assets on demand.

  Trade execution in AMM protocols is performed via liquidity pools for each pair of tradable tokens, reserved in their respective smart contracts. A trader looking to exchange $X$ tokens for $Y$ tokens, can deposit $X$ tokens in the liquidity pool, and receive $Y$ tokens in an atomic swap, such that the aggregate liquidity of the pool remains unchanged, as defined by the bonding curve (Park, 2022). This pricing function determines the exchange rate to swap the tokens.

  There exist a class of traders known as arbitrageurs. These arbitrageurs identify pricing differentials of assets that exist between different exchanges, and trade such differentials to extract profit. When the exchange rate of a token pair deviates from other exchange quoted prices, AMM protocols allow arbitrageurs to execute arbitrage trades, so as to bring the exchange rate closer to general market conditions (Aoyagi and Ito, 2021).

- *Liquidity provider*:

  A liquidity provider is any party that contributes liquidity to the market. They create an efficient market where liquidity takers can transact assets.

  Liquidity providers commit pairs of $X$ and $Y$ crypto-assets to the pool, so that liquidity exist for traders to buy and/or sell $X$ and $Y$ crypto-assets. Liquidity providers are incentivized through market-making incentive fees from the trades supported by their liquidity.

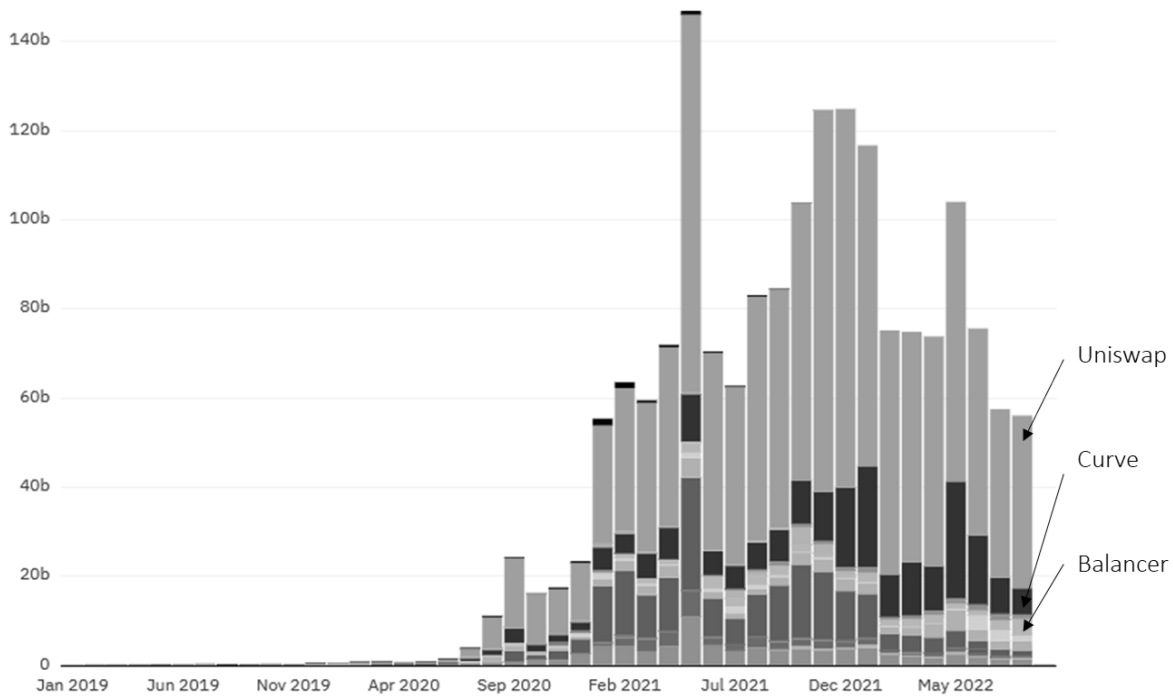

**Figure 1. Monthly DEX transaction volume by project (Dune Analytics, 2022)**

From a monthly trading volume perspective, at the time of writing, at the month of August 2022, Uniswap leads the AMM market by a distance, with 39 billion trades, outstripping the next two highest AMMs Curve and Balancer, at 6 and 2 billion respectively (Figure 1). At its peak, there were 86 billion trades traded on Uniswap. Other popular protocols are Sushiswap, Synthetix, DODO and 0x Native.

Most key AMMs on Ethereum-based protocols implement a constant function market maker (CFMM) for execution of trades (Uniswap, 2022; Curve, 2022; Balancer, 2022, Sushiswap, 2022). CFMM are AMMs that utilize a fixed bonding curve for asset price determination and liquidity provision. Angeris and Chitra (2020) showed that agents interacting with CFMMs are incentivized to price assets correctly, in computationally efficient manners.

In this paper, we focus on the top utilized protocol – Uniswap. Uniswap has two actively traded versions V2 and V3. Uniswap implements the $XYC$ constant product market maker (CPMM) function, where given $x$ units of token $X$ and $y$ units of token $Y$, the liquidity of the pool $K$ is the product of $x \cdot y = c$. Upon choosing a pool to provide liquidity, Uniswap V2 allows a liquidity provider to supply liquidity across the entire price range, whereas Uniswap V3 applies a novel CPMM design that allows liquidity providers to specify the price range at which they wish to supply liquidity. Since introduction, Uniswap V3 has overtaken V2 to become the AMM with the largest trading volume.

However, despite the high trading volumes, there continues to exist issues for both the liquidity pool and market participants in Uniswap V3.

- *Liquidity pool*:

    Capital efficiency is a function of the amount of capital needed to provide for an efficient market making. The less capital required to make the market, the more

efficient is the liquidity provision. This also implies total value locked (TVL) is not a useful metric to measure liquidity productiveness of a liquidity pool.

- *Liquidity taker:*

    Slippage is an implicit cost to a liquidity provider that occurs when the price at which a trade is executed, and the expected price of the trade, are different. Slippage can occur, when the market is volatile, or when the sizes of the trades are large relative to the size of the liquidity pool. While slippage cannot be entirely eliminated, it is to the benefit of liquidity takers to lower this market inefficiency to the lowest possible.

- *Liquidity provider:*

    Allowing selection of price range to supply liquidity changes the risk-return dynamics of liquidity provision to the liquidity provider, such that users who select the right price position and width to concentrate liquidity will be well rewarded to counter the effect of divergence loss, in contrast to those who do not.

    Divergence loss (or impermanent loss) is an implicit cost to a liquidity provider tied to the risk of a decline in value of the liquidity position, when compared to the value of the initial deposited assets. Heimbach, Schertenleib & Wattenhofer (2022) demonstrated how liquidity providers' risk-return profile of selected liquidity ranges in Uniswap V3 can show significant fluctuations, which may require active management strategies to circumnavigate. Further, such active managing of positions can affect market depth in volatile market conditions, counter to the interest of the AMM protocol.

The study proposes a quote-driven AMM with its original intent of on-chain custody and settlement functions, alongside off-chain predictive reinforcement learning capabilities. Firstly, the proposed AMM architecture is an augmentation to the Uniswap V3 protocol by utilizing a novel market equilibrium pricing for reduced divergence and slippage loss. Secondly, the proposed protocol involves a predictive automated market making capability, utilizing a deep hybrid reinforcement learning framework that looks to improve market efficiency through better forecasts of liquidity concentration ranges, so liquidity starts moving to expected concentration ranges, prior to asset price movement so that liquidity utilization is improved.

The augmented protocol framework is expected to (i) reduce divergence loss for liquidity providers, and (ii) reduce slippage for crypto-asset traders, while (iii) improving capital efficiency for liquidity provision for AMM protocol. To our best knowledge, there are no known protocol or literature that are proposing similar deep learning-augmented AMM that achieves similar capital efficiency and loss minimization objectives.

**Related Work**

*Pricing and Loss in AMM DEX*

Xu et al. (2022) discussed the economics of AMM DEX, including rewards such as liquidity incentive fees, and implicit costs such as divergence and slippage losses. Heimbach, Schertenleib & Wattenhofer (2022) analyzed factors influencing the performance of liquidity positions in Uniswap V3, including divergence loss and selection of liquidity positions. Aoyagi

(2020) proposed the application of an equilibrium valuation point for more accurate pricing in AMM DEX. Engel and Herlihy (2021b) provided a good analysis on how equilibrium valuation price and divergence and slippage losses can be minimized in AMM DEX, based upon the formal model, axioms and notations in Engel and Herlihy (2021a). Engel and Herlihy (2021a; 2021b) provides the foundational work of this paper.

*Deep Reinforcement Learning on AMM DEX*

Application of reinforcement learning on market making started as early as Chan & Shelton (2001). Most recently, Hambly, Xu & Yang (2021) provided an account of the state-of-the-art of reinforcement learning on market making.

Market making is generally applied in market microstructure modelling research using the stochastic control or reinforcement learning approaches, where optimal bidding, such as pricing strategy in limit order books (LOB), is studied (Sun, Huang & Yu, 2022). This study restricts the focus to the application of reinforcement learning on AMM DEXs which operates in an algorithmically deterministic market making manner, rather than LOB. Pourpouneh, Nielsen & Ross (2020) provided a survey of present AMM models.

Research on this sub-domain is sparse. Most crypto-asset-based research applying deep reinforcement learning, is in relation to automated trading from an investment management perspective, covered to some extent by Lucarelli & Borrotti (2019). In relation to DEX, Sadighian (2019; 2020) proposed, and later enhanced, a deep reinforcement learning framework for a crypto-asset DEX, using a policy gradient-based algorithm to interact with data from limit order book and order flow arrival statistics to solve a stochastic inventory control problem. There is limited research that applies deep reinforcement learning on crypto-asset-based AMM DEX.

**Proposed Method**

*Notation*

We define the notations in this paper, as per Engel and Herlihy (2021a). Italics is used for scalars ($x$) and bold typography for vector ($\boldsymbol{x}$). Constants are defined from the beginning of the alphabet ($a, b, c$), and variables, vectors, or scalars from the end ($x, y, z$). We represent "=" for equality and ":=" for definitions. We represent subscript "$obs$" as a market observed price and subscript "$p$" as predicted valuation.

Informally, to represent CPMM, an AMM in state $(x, y)$ holds the custody of $x$ units of token $X$ and $y$ units of token $Y$, subjected to $x \cdot y = c$, where $x, y > 0$ and some constant $c > 0$. For any trade to occur, liquidity invariance is achieved when a buyer purchasing $\delta_X$ of token $X$, will deposit $\delta_Y$ of token $Y$, such that $(x - \delta_X) \cdot (y + \delta_Y) = c$.

To formally represent CPMM, an AMM state trading assets $X$ and $Y$ is represented by $(x, y) \epsilon \mathbb{R}^2_{>0}$. The state space is represented by curve $(x, f(x))$, such that $f: \mathbb{R}_{>0} \to \mathbb{R}_{>0}$. It is assumed that the pool of assets is not exhausted and boundary conditions are set as $\lim_{x \to 0} f(x) = \infty$ and $\lim_{x \to \infty} f(x) = 0$.

It is noted that Uniswap charges fees of 0.3% for each trade back to the asset pool, which in part is used to incentivize the liquidity providers. Here the effect of this fee is ignored, as they have minimal impact on costs. In general, fees will cause a slight reduction in divergence loss for liquidity providers and slippage cost for liquidity takers.

### *Equilibrium State*

For asset pricing, it is assumed that only one market valuation is acceptable to most liquidity takers at any time. *Valuation* $v \in (0,1)$ is assigned, such that $v$ units worth of $X$ equates to $(1-v)$ units worth of $Y$. At valuation $v$, a profit is made when $v(x - x') + (1-v)(f(x) - f(x'))$ is positive when the AMM state space moves from $(x, f(x))$ to $(x', f(x'))$. Otherwise, a loss is incurred.

An *equilibrium point*, or the state at which no arbitrage profits can be made, is defined to be a valuation $v$ at point $(x, f(x))$ that solves the optimization problem (Equation 1):

$$v := \min_x v_{obs} \cdot x \tag{1}$$

where $v_{obs}$ is the market observed price on the asset obtained from a trusted price oracle.

For $(x, f(x))$ to be the equilibrium point, $\frac{df(x)}{dx} = -\frac{v}{1-v}$. The exchange rate of asset $Y$ in units of asset $X$ is defined as $-f'(x)$. This is the negative of the curve's slope at the point.

To carry each valuation $v$ to the equilibrium state $x$ that minimizes the dot product $v_{obs} \cdot x$, or $vx + (1-v)f(x)$, we define $\phi(v) = f'^{-1}(-\frac{v}{1-v})$, where $\phi: (0,1) \to \mathbb{R}_{>0}$. For instance, the equilibrium state for AMM at $\left(x, \frac{1}{x}\right)$ is $\phi(v) = \sqrt{\frac{1-v}{v}}$. It is useful to express $\phi$ in vector representation $\mathbf{\Phi}(v) := (\phi(v), f(\phi(v)))$, where $\mathbf{\Phi}: (0,1) \to \mathbb{R}^2_{>0}$. The inverse of $\phi$ is represented by $\psi(x) = -\frac{f'(x)}{1-f'(x)}$, where $\psi: \mathbb{R}_{>0} \to (0,1)$. The vector representation is $\mathbf{\Psi}(x) := (\psi(x), 1 - \psi(x))$.

It is noted that every $x$ is the equilibrium point for some valuation $v$. For instance, for a CPMM AMM $:= \left(x, \frac{1}{x}\right)$, the point $\left(x, \frac{1}{x}\right)$ is the equilibrium point for $\left(\frac{1}{1+x^2}, 1 - \frac{1}{1+x^2}\right)$. To generalise, for a CPMM AMM $:= (x, f(x))$, the point $(x, f(x))$ is the equilibrium point for $\left(\frac{f'(x)}{f'(x)-1}, \frac{f'(x)}{1-f'(x)}\right)$ (Engel and Herlihy, 2021a).

### *Total Value of AMM Holdings*

Let the valuation with equilibrium point $(x, f(x))$ be defined as $(v, 1-v)$. Given $\boldsymbol{v} = (v, 1-v)$ and $\boldsymbol{x} = (x, f(x))$, the total value (or *capitalization*) of the total AMM holding is given by (Engel and Herlihy, 2021b):

$$cap(x, v) := vx + (1-v)f(x) = \boldsymbol{v} \cdot \boldsymbol{x}$$

In an event when $v$ represents the current market valuation, the AMM is in the equilibrium state $\Phi(v) = (\phi(v), f(\phi(v)))$, giving:

$$cap(v) := cap(\phi(v), v) = v \cdot \Phi(v)$$

In the case of a CPMM AMM $:= \left(x, \frac{1}{x}\right)$, the capitalization at equilibrium point is given by:

$$cap\left(v; \left(x, \frac{1}{x}\right)\right) := 2\sqrt{v(1-v)}$$

### *Divergence Loss, Slippage Loss and Load*

To improve the performance of an AMM utilizing CPMM function, we look to reduce divergence and slippage losses. This section defines divergence and slippage losses (Engel and Herlihy, 2021b), and identify a composite loss function to reduce these losses.

- *Divergence loss:*

    *Divergence loss* is incurred when there is a difference in value arising from the funds remaining in the wallet, against the initial fund amount deposited into the AMM. In an event the valuation $v$ moves to $v'$, the equilibrium state shifts from $x$ to $x'$. The shift away from $v$ creates an unstable state, such that arbitrageurs will be able to profit the amount of $v' \cdot x - v' \cdot x'$.

    Divergence loss is defined as a function of liquidity pool size, as follows:

    $$\begin{aligned} loss_{div}(v, v') &:= v' \cdot \Phi(v) - v' \cdot \Phi(v') \\ &= v'\phi(v) + (1-v')f(\phi(v)) - (v'\phi(v') + (1-v')f(\phi(v'))) \end{aligned}$$

    where $\Phi(v, 1-v) = (\phi(v), f(\phi(v)))$.

    In the case of a CPMM AMM $:= \left(x, \frac{1}{x}\right)$, divergence loss for trade size $\delta$, is given by:

    $$loss_{div}(x, x+\delta) := \frac{\delta^2}{2\delta x^2 + x^3 + \delta^2 x + x}$$

- *Slippage loss:*

    *Slippage* loss is defined by how an increase in trade sizes can reduce a liquidity taker's return. Suppose a trade size of $\delta$ is placed, where $\delta > 0$. The state of the AMM changes from $(x, f(x))$ to $(x+\delta, f(x+\delta))$. In a linear rate of exchange, in exchange of $\delta$ units of $X$, the trader receives $-\delta f'(x)$ units of $Y$. Therefore, the trader makes a loss of $-\delta f'(x) - f(x) + f(x+\delta)$, resulting in the final receipt of $f(x) - f(x+\delta)$.

    Slippage is defined as a function of liquidity pool size, as follows:

    $$loss_{slip}(v, v') := \left(\frac{1-v'}{1-v}\right)(v \cdot \Phi(v') - v \cdot \Phi(v))$$

In the case of a CPMM AMM $:= \left(x, \frac{1}{x}\right)$, divergence loss for trade size $\delta$, is given by:

$$loss_{slip}(x, x + \delta) := -\frac{\delta^2(\delta+x)}{x^2(\delta^2+x^2+2\delta x+1)}$$

- *Composite divergence and slippage loss:*

  To reduce the overall effect of cost of divergence loss to liquidity providers and slippage loss to liquidity takers, a composite function, known as *load* (Engel and Herlihy, 2021b), taking into account both divergence and slippage losses can be useful. Load across an interval, with respect to $X$, is defined as the product of interval's slippage and divergence loss, given by:

  $$load_X(v, v') := loss_{div}(v, v') \cdot loss_{slip}(v, v')$$

  Given a probability density for future valuations, we can compute an *expected load* when exchanging $X$ tokens for $Y$ tokens, starting in the equilibrium state for valuation $v$. Given $p(v')$ is the distribution over possible future valuations (Equation 2):

  $$E_p[load(v')] := \int_0^v p(v') load_X(v, v') dv' + \int_v^1 p(v') load_Y(v, v') dv' \quad (2)$$

***Pricing and Changes to Liquidity Provision***

Suppose for an AMM $:= (x, f(x))$, the valuation moves from $v$ with equilibrium state $(a, b)$, to $v'$ with equilibrium state $(a', b')$. An arbitrageur can make an arbitrage profit by moving from $(a, b)$ to $(a', b')$.

We can eliminate this arbitrage that results in divergence loss, by moving the bonding curve in the AMM protocol, as a *pseudo arbitrage*, as referred to by Engel and Herlihy (2021b). Suppose $a > a'$ and $b' > b$, the transformed AMM becomes $AMM' := \left(x, f(x - (a - a')) - (b' - b)\right)$. The new equilibrium state $v'$ = new market price, and continues to lie on the shifted curve with a slope of $\frac{v'}{v'-1}$.

A downside of this above illustrated pseudo arbitrage is that the AMM now has more units of $X$ and a shortage of $Y$ to cover all possible trades (Engel and Herlihy, 2021b). This imbalance is small, as each price action is generally driven by small tick changes, assuming an efficient market. However, they can add up over time and become problematic. As a result, the AMM will have to account for this shortfall, by making minor adjustments to liquidity provision to rebalance the pool. This implies that, as part of liquidity provision, liquidity providers will deposit an additional $X$ or $Y$ tokens as stated by the AMM. Incentives will be given for all tokens deposited, including the additional $X$ or $Y$ tokens. It is proposed to incorporate this in the configurable virtual AMM, as shown in the proposed AMM architecture in a later section. In this way, the bonding curve will revert back to its primary CPMM bonding curve formula.

***Deep Reinforcement Learning***

Where there exists an interaction between the agent (AMM) and the environment (financial market, including market participants such as liquidity takers and providers), we can execute actions and receive observations and rewards as a Markov Decision Process. At each time step $t$, the agent selections an action $a_t \in \mathcal{A}$ at state $s_t \in \mathcal{S}$, where $\mathcal{S}$ is the set of possible states. The step of action selection depends on the policy $\pi$, which is a description of the agent behaviour, and it guides the actions the agent takes for each possible state. Upon the execution of each action, a scalar reward $r_t \in \mathcal{R}$ is received by the agent, and the next state $s_{t+1}$ is observed. This learning sequence will be repeated in a (possibly infinite) horizon $T$, until the algorithm is halted. The transition probability of possible future state $s_{t+1}$ is given by $P(s_{t+1}|s_t, a_t)$, and the reward probability is given by $P(r_t|s_t, a_t)$. Therefore the expected reward is computed as $E_{P(r_t|s_t, a_t)}(r_t|s_t = s, a_t = a)$.

- *Event-driven environment*

    This study considers a state-based AMM agent that acts on events as they occur. The action-space is based on a typical market making strategy where the agent cannot exit the market and is restricted to executing a single order. An event constitutes an observable change in the state of the environment and can occur due to a change in price. This implies that actions are not regularly spaced in time. The agent is required to quote prices which it is willing to buy and sell at valid time points, unless constraints to asset inventory prevail.

    In line with Sadighian (2020), the study proposes the use of price-based approach for event-driven environment, where an event is defined as change in equilibrium valuation $v'$, and if this is greater than or less than a threshold $\beta_v$. $\beta_v$ allows the adjustment of the sensitivity of the rate of learning.

    These price change events are not regularly spaced in time, which reduces the time required to train the agent per episode (ie. an executed trading action resulting in price change). Algorithm 1 shows the algorithm to evaluate price-based event (Sadighian, 2020).

| **Algorithm 1: Evaluating Price-based Event** |
|---|
| **Input:** Valuation $v'_t$ at time $t_0 = 0$ <br> **Output:** Observation at time $t_k = t_0 + k$ <br><br> 1    $\beta_v \leftarrow 0.01\%$ <br> 2    $k \leftarrow 0$ <br> 3    $upper \leftarrow v'_t(1 + \beta_v)$ <br> 4    $lower \leftarrow v'_t(1 - \beta_v)$ <br> 5    $step \leftarrow True$ <br> 6    **while** $step$ **do** <br> 7      **if** $upper \leq v'_{t+k} \leq lower$ **then** <br> 8        $k \leftarrow k + 1$ <br> 9      **else** <br> 10       $step \leftarrow False$ <br> 11      **end** <br> 12  **end** |

- *Reward function*

  We recognize that the obvious reward functions in most state-of-the-art reinforcement learning for market making literature select profit seeking (Spooner et al., 2018; Sadighian, 2020; Haider et al, 2022), or utility maximization (Selser, Kreiner & Maurette, 2021) as the natural choices of reward functions.

  In this study, to improve market efficiency and provide optimal liquidity, the reward objective function for trading agents is tied to the quality of forward prediction of valuation $v'_p$, against the equilibrium valuation $v'$ at this future time, and implicit costs for liquidity takers and providers.

  We propose a single-step loss function $\ell$ as follows (Equation 3):

  $$\ell := |v'_t - v'_{p,t}| + E_p[load(v')] \tag{3}$$

  This loss function (Equation 3) computes the *prediction slippage*, or the difference between the valuation $v'_p$ as predicted by an AMM prediction model, against the equilibrium point $v'$ (computed from Equation 1). The latter is a function of the actual observed valuation from a trusted price oracle. We take the modulus of this difference as we want to identify absolute deviations between prediction and equilibrium prices, so as to minimize this difference using reinforcement learning. Further, we add the expected load (computed from Equation 2), which represents the divergence and slippage losses. The overall objective is to minimize this function, by reducing prediction slippage and expected load, and in turn, improves capital efficiency.

  The cumulative reward function $R$ as follows:

  $$R_t := \sum_{k=0}^{k=T} \gamma^k r_{t+k}$$

  where $\gamma \in (0,1)$ is a parameter called the discount rate. $r$ is defined as:

  $$r_{t+k} := \begin{cases} -1, & if\ \ell_t > \beta_c \\ 0, & if\ \ell_t = \beta_c \\ +1, & if\ \ell_t < \beta_c \end{cases}$$

  where $\beta_c$ represents a threshold within which prediction slippage and expected load can be tolerated. This threshold determines the sensitivity of the reward function to the loss function (Equation 3).

- *Action space*

  The agent action space consists of 2 possible actions:

  $$A_t := \begin{cases} Insert\ input\ parameter\ \varepsilon_{t+k}, & if\ \ell_{k-t} > \beta_c \\ Do\ nothing, & if\ \ell_t \leq \beta_c \end{cases}$$

  where $\varepsilon$ represents a gaussian input parameter to the learning model, where $\varepsilon \in (-1,1)$ and $\varepsilon \sim N(\mu_\varepsilon, \sigma_\varepsilon)$. This input parameter effects changes to the learning model, with the goal to help reduce prediction slippage and expected load.

- *State space observations*

An environment state is constructed from an attribute set that describes the condition of the market and agent.

The market state comprises observations derived from, among others:

- Market valuation obtained from external trusted price oracle, represented by $v_{obs}$.

- Pre-processed alternative data that indicate price signals that effect changes in market liquidity, represented by $\tau$.

    An example of such alternative data sources is market signals generated from Twitter data processed using natural language processing to make predictions. (Abraham et al., 2018; Kraaijeveld & De Smedt, 2020). In this paper, we pretrain a Long Short-Term Memory (LTSM) supervised learning model, and utilize the LSTM outputs as observation inputs for reinforcement learning (Liu, 2020).

The agent state comprises observations derived from trading agent's own records, including, among others:

- Number of units of token $X$, represented by $x$, and number of units of token $Y$, represented by $y$

- *Q-learning*

    The expected discounted return at time $t$ is defined as $R_t \coloneqq E[\sum_{k=t}^{k=T} \gamma^{k-t} r_{k-t+1}]$. Applying Q-learning as a recursive update procedure, the Q-value function $Q^\pi(s, a)$ is defined as:

    $$Q_{i+1}^\pi(s, a) \coloneqq E_\pi[r_t + \gamma \sum_{k=0}^{k=T} \gamma^k r_{t+k+1} | s_t = s, a_t = a]$$
    $$= E_\pi[r_t + \gamma Q_i^\pi(s_{t+1} = s', a_{t+1} = a') | s_t = s, a_t = a]$$

    Reinforcement learning learns the optimal policy $\pi^*$ whose expected value is greater than or equal to all other policies, to converge at an optimal Q-value $Q^*(s, a)$.

    $$Q_{i+1}(s, a) \coloneqq E_\pi[r_t + \gamma \max_{a' \epsilon A} Q_i(s', a') | s, a]$$
    $$Q^*(s, a) \coloneqq (\mathcal{B}Q^*)(s, a)$$

    where $\mathcal{B}$ represents the Bellman operator that maps any function $\mathcal{K}: S \times A \mapsto R$ into another function $S \times A \mapsto R$. Bellman operator is given as follows:

    $$(\mathcal{B}\mathcal{K})(s, a) \coloneqq \sum_{s' \epsilon S} \mathcal{T}(s, a, s') [R(s, a, s') + \gamma \max_{a' \epsilon A} K(s', a')]$$

    where $\mathcal{T}$ represents the function to compute the transaction value to move from $s$ to $s'$, given an action $a$.

- *Deep reinforcement learning architecture*

    This paper proposes a hybrid LSTM-Q-learning architecture, with its architectural derivatives proposed in Lucarelli & Borrotti (2019) and Liu (2020).

To perform prediction for the forward valuation $v'_p$, Liu (2020) found usefulness to perform pretraining of a supervised recurrent neural network in the form of Long Short-Term Memory (LTSM) and utilize the LSTM outputs as observation inputs for reinforcement learning. 1 LTSM layer is applied with 100 neurons, with sliding window of 50 interval inputs including (i) the market observed price from the trusted oracle $v_{obs}$, (ii) pre-processed alternative data representing market movement signals $\tau$, and (iii) a gaussian input parameter from the action space $\varepsilon$ that seeks to reduce prediction slippage and load.

The predicted output of $v'_p$, computed equilibrium price $v'$, and computed load $E_p[load(v')]$ are used as inputs for the Q-learning model. For the D-DQN architecture, 2 CNN layers are applied with 100 neurons each. In the case of DD-DQN architecture, 2 CNN layers are applied each with 100 neurons, followed by two fully connect layer streams – one with 50 neurons used to estimate the value function, and another with 50 neurons to estimate the advantage function. Both epochs and batch sizes are set to 50. Weight optimization applies Adam algorithm (Kingma & Ba, 2015). Activation function applies Leaky Rectified Linear Units (Leaky ReLU) (Maas, Hannun, & Ng, 2013). $\gamma$ is set to 0.98 (Lucarelli & Borrotti, 2019). Given Equation 3, the loss function is defined as follows (Equation 4).

$$\mathcal{L} := \frac{1}{n}\sum_{k=0}^{k=T} \{ |v'_t - v'_{p,t}| + E_p[load(v')] \} \qquad (4)$$

The proposed recursive LSTM-Q-learning DD-DQN reinforcement learning architecture is shown in Figure 2. Details of neural network layers are shown in Figure 3.

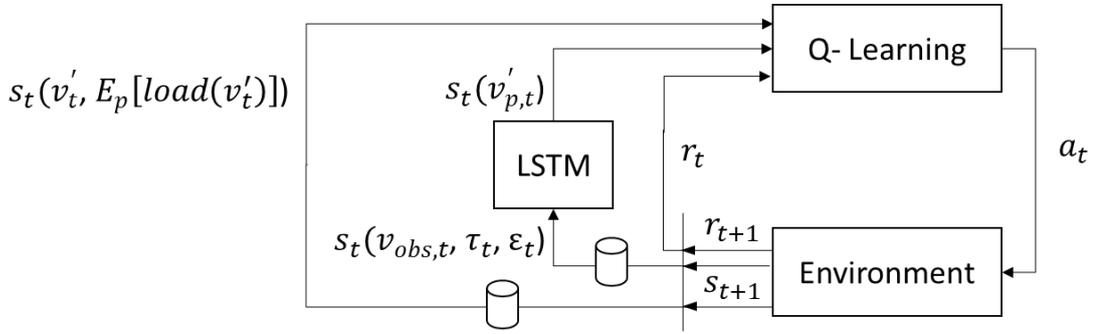

**Figure 2. Recursive LSTM-Q-learning reinforcement learning architecture**

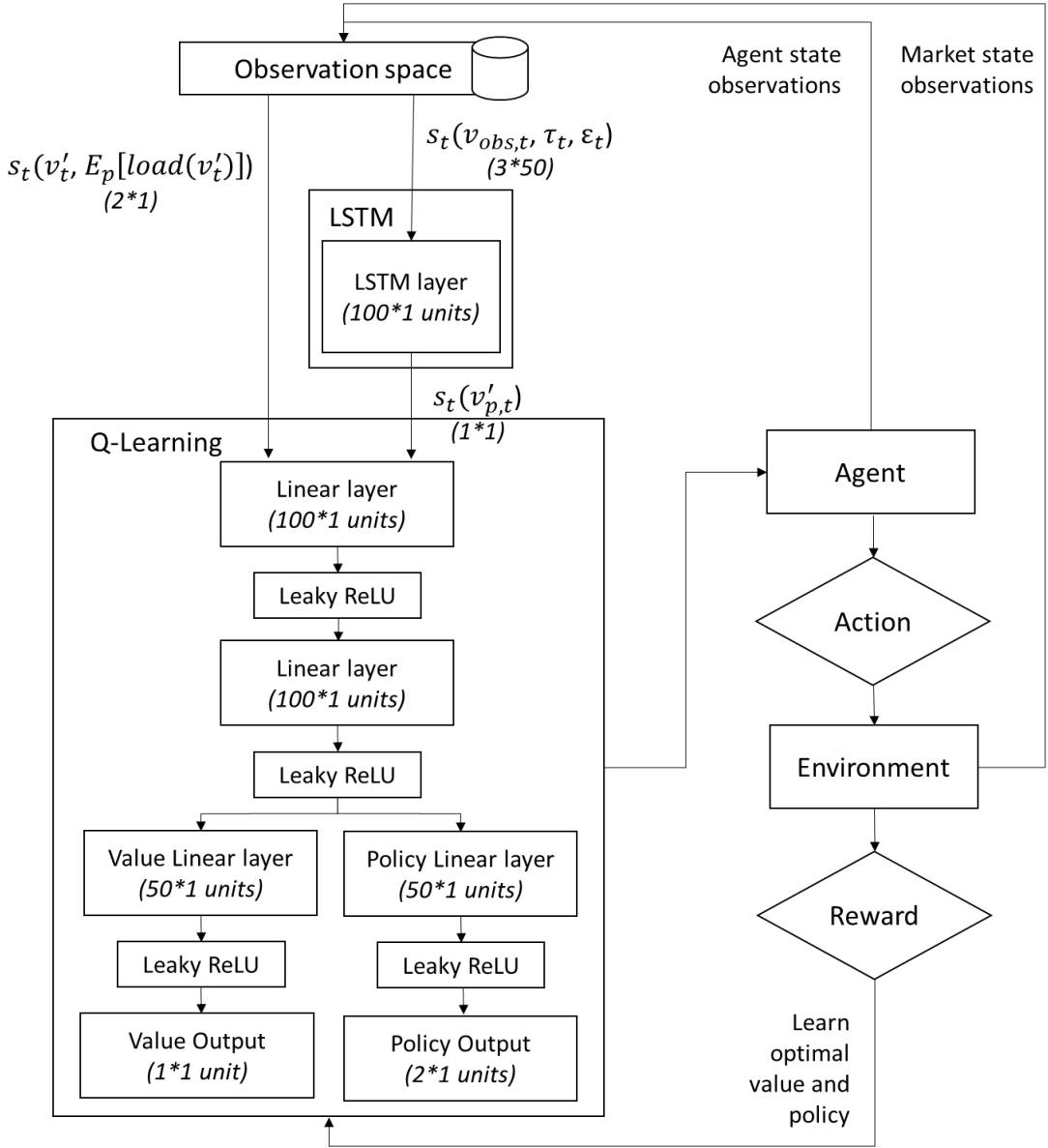

**Figure 3. LSTM-Q-learning reinforcement learning architecture layers**

*Predictive Liquidity Distribution*

Present liquidity concentration ranges where incentives to liquidity providers are distributed, are done on a "look back" basis, relying on observed market values (Uniswap, 2022). A market maker is responsible for providing liquidity for trade execution. Liquidity pooling requires time to form. Through advanced prediction of valuation $v'_p$, incentivization for liquidity provision can be altered in $n$ intervals in advance (*e.g.* 1, 5 or 10 intervals), so liquidity shifts prior to the actual market change. The shifting of incentive fee distribution can help motivate liquidity providers in seek of higher yields to support predicted new liquidity concentration ranges, so that pool capital efficiency can be achieved.

Further, current incentivization program for Uniswap V3 liquidity providers is binary in nature, such that fees will only be earned if liquidity providers provide liquidity within a certain range in the bonding curve, and will not be compensated if their liquidity provision falls outside the range. However, research has shown that the proportion of time where asset prices remain within a liquidity position relative to a liquidity width is not a uniform distribution (Heimbach, Schertenleib & Wattenhofer, 2022). While active liquidity providers benefit from range targeting to earn the best possible fees in a uniform distribution fee structure in Uniswap V3, it is useful to consider a different distribution structure that can help "insure" against sharp price movements, which can help improve the attractiveness of liquidity provision.

We use $v'_p$ to help determine the position of the new liquidity concentration range on the bonding curve. The distribution of incentive fee $\varphi$ is proposed to be gaussian in nature (Figure 4), such that $\varphi \sim N(\mu v_\varphi, \sigma_\varphi)$ and $\mu_\varphi = v'_p$, and is given by:

$$\varphi(x) := \frac{1}{\sigma_\varphi \sqrt{2\pi}} e^{-\frac{1}{2}(\frac{x-v'_p}{\sigma_\varphi})^2}$$

In effect, the LSTM-predicted $v'_p$ formulates the new liquidity concentration region on the bonding curve in $n$ intervals in advance, so the liquidity pool rebalances its liquidity before actual market change occurs. Active yield seekers who shift liquidity to new liquidity concentration positions will be rewarded positively. Further, as incentivization distribution is gaussian in nature, liquidity providers continue to be incentivized, albeit to lesser amounts, even if they do not correctly identify the best prices and length of time to position their liquidity provision. As compared to Uniswap V3, this relative lowering of "incentive penalization" due to incorrect liquidity positioning, looks to help draw liquidity providers.

For this purpose, it is also proposed to include transparency in $v'_p$, and the historical shifts in $v'_p$ in the AMM design to liquidity providers, so as to positively improve market's ability to analyze and pre-position resource allocation.

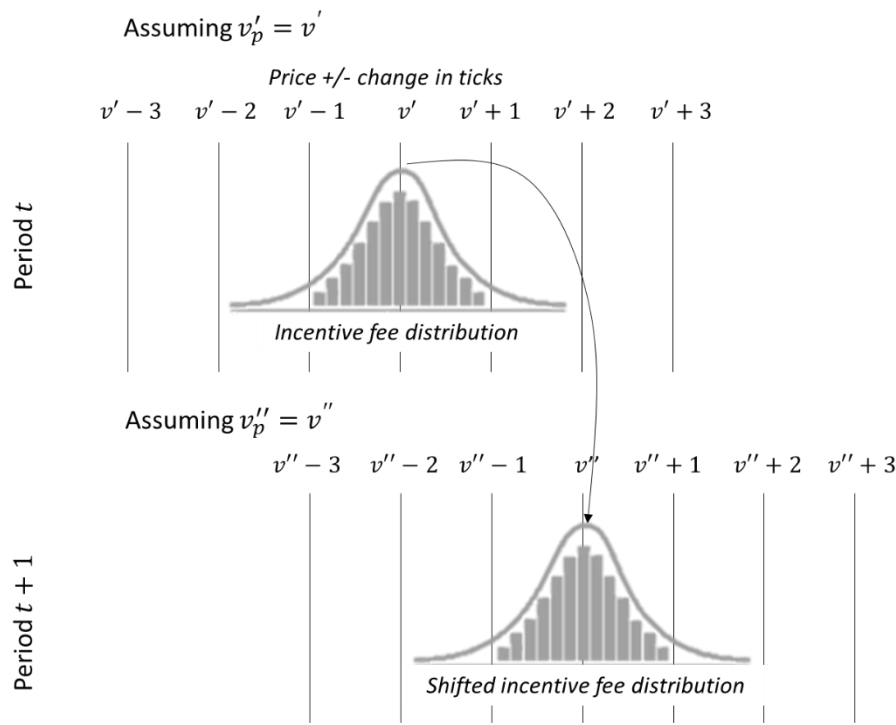

**Figure 4. Incentive distribution for liquidity providers**

*Proposed Architecture*

The proposed architecture in Figure 5 comprises the following protocol layers (Xu et al., 2021), augmenting deployment from Shrivastava (2022). Interactions between key architecture layer components are summarized in Figure 6.

- *Aggregator layer*

    This layer extends the application layer, designed to create user-centric platforms to connect several protocols and applications, so that users can connect to multiple protocols, and perform tasks, such as transact across services and compare services.

- *Application layer*

    The application layer comprises 2 components, the blockchain layer interaction service and the user interface.

    - *User interface:* This is designed to allow AMM users to interact with the various system functions provided by the AMM. This is usually abstracted by a web browser or mobile application-based front end.

    - *Blockchain layer interaction service:* This is designed to communicate and interact with the smart contract protocol layer. The interaction service allows function calls to be applied in the clearing house to perform specific actions a liquidity provider or taker carries out.

- *Blockchain protocol layer*

    This is the protocol layer for asset pooling and transaction settlement. The system logic is contained in the smart contracts deployed on a blockchain (e.g. Ethereum blockchain). A layer 2 solution may be implemented to allow off-chain transactions that can be rolled up to the layer 1 Ethereum main chain, to lower Ethereum gas fees and improve processing rate.

    - *Clearing house:* This is designed to securely execute trade positions, and facilitate the deposit and returning of funds when called upon by a liquidity provider or taker. It is also responsible for the returning of details about the vault, price of the token pair, and token reserves in the AMM.

    - *Configurable virtual AMM (cAMM)*: This protocol allows the flexibility to adjust token pair price $v'$ based on spot market prices $v_{obs}$ to minimize expected load. Liquidity is also recalibrated based on $v'_p$ for the determination of the liquidity concentration range and the distribution of incentive fee $\varphi$.

    - *Vault*: This smart contract vault holds the deposits securely for liquidity providers and takers.

    - *Oracle*: This protocol allows the discovery of spot price for a token pair.

- *Infrastructure layer*

This layer contains the trusted execution environment (TEE), which provides an enclave for secure intensive computation, such as the proposed LSTM-Q-learning reinforcement learning model, where external applications outside the enclave will not be able to interfere with the state or control flow shielded by the TEE (Pandl et al., 2020).

- *TEE:* This physical server environment is designed to enable computation of resource intensive machine learning applications, while preserving data integrity and security throughout the compute process. Through smart contracts, protocols can be designed to define policies on how data is shared. The policies may include the requests for reward and differential privacy requirements (Hynes, Cheng & Song, 2018). As the deep reinforcement learning model is shielded by the smart contract and inference executions count towards the contract policies, this improves privacy against potential inference attacks, which aims to execute the predictive system to extract the model or underlying data (Cheng et al., 2019).

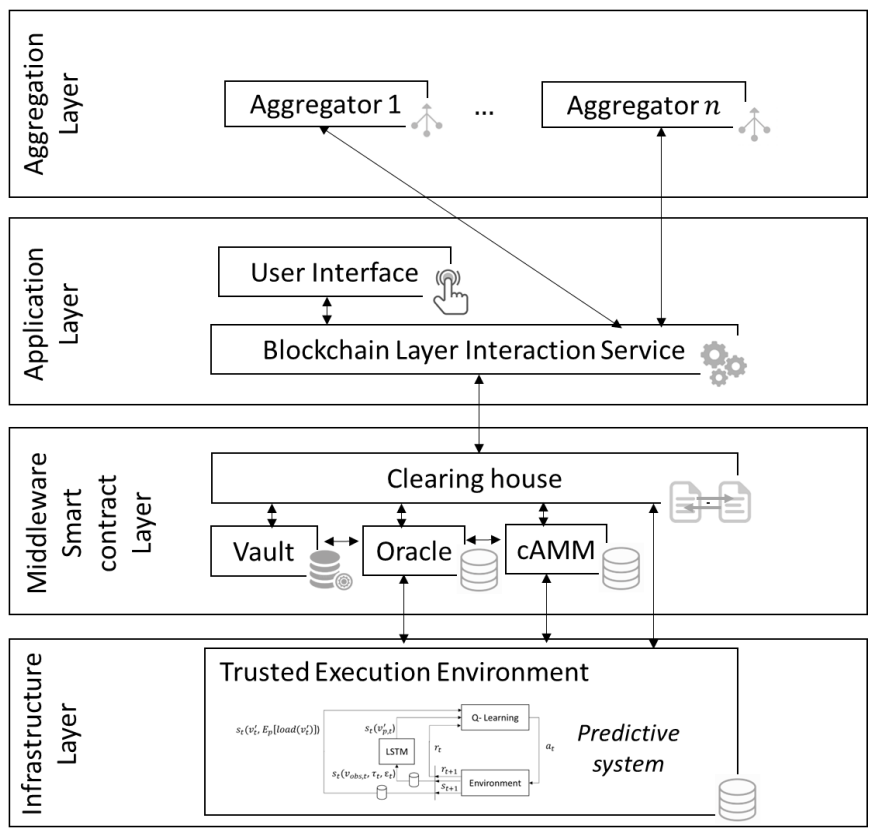

**Figure 5. Architecture layers of proposed AMM**

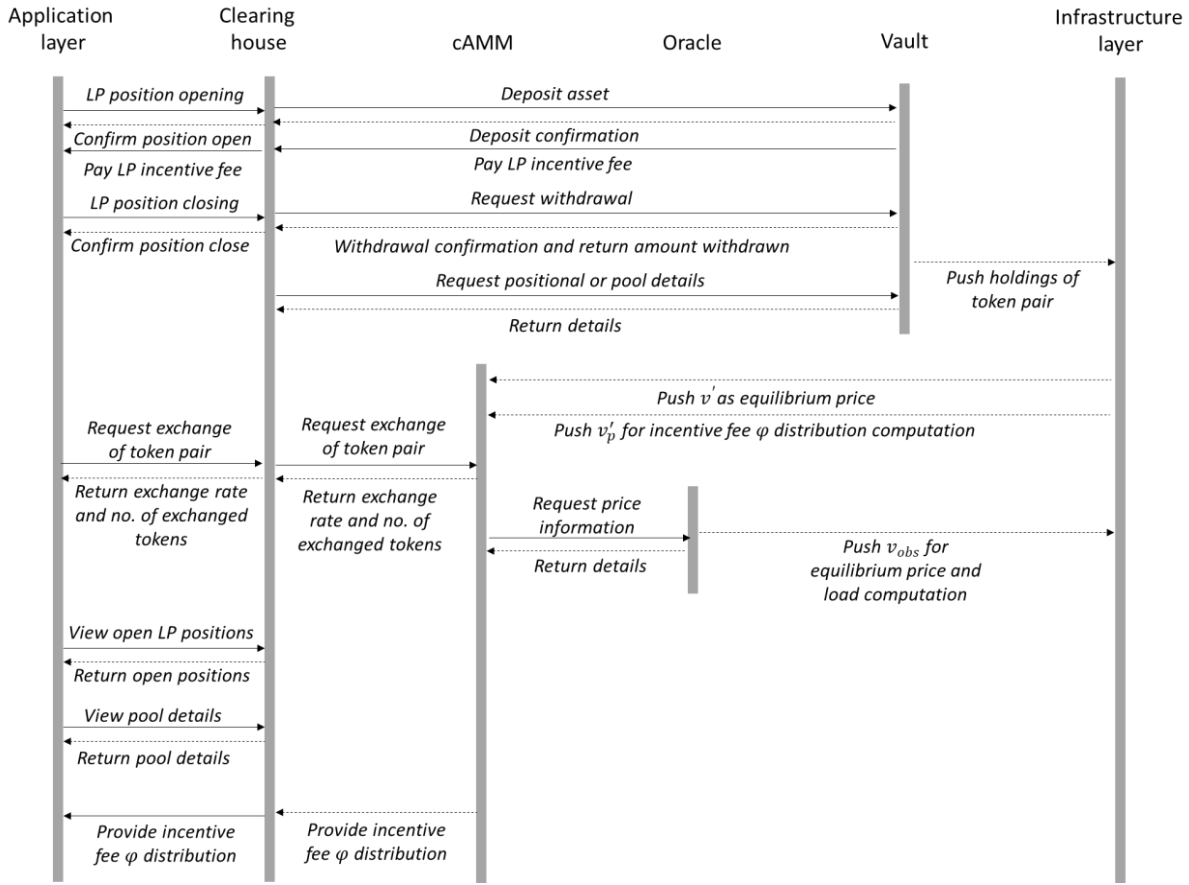

Figure 6. Interaction between key architecture layer components of proposed AMM

**Conclusion**

AMM DEX is a recent development in its early stages of growth. Present market solutions, while innovative in nature, can be further optimized.

In this work, a predictive AMM architecture is introduced, that utilizes a loss-minimizing market pricing mechanism, and a deep reinforcement learning architecture that looks to reduce divergence and slippage costs, with the objective of enhancing capital efficiency of liquidity provision. The paper formalizes and analytically exposits the implicit costs to a liquidity taker and provider, and the deep reinforcement learning mechanism for market making, to benefit research and industry use.

An AMM DEX liquidity provision optimization strategy is an attractive topic for both practitioners and researchers. For practitioners, future development work can look to include a physical implementation of the proposed AMM DEX, built upon a profitable business model. For researchers, proposed further research can include: (i) an evaluation of a range of pre-processed alternative data that indicate price signals effecting changes in market liquidity, (ii) an evaluation of incentive fee distribution structures beyond the Uniswap V3 uniform distribution and the paper-proposed gaussian distribution mechanisms, (iii) an investigation of an integrated TEE architecture within the infrastructure layer, augmented with relevant specific security protocols, and (iv) an improvement on the proposed hybrid LSTM-Q-learning reinforcement learning framework to enhance prediction of liquidity concentration ranges.

## Statements and Declarations

There are no financial or non-financial interests that are directly or indirectly related to the work submitted for publication.